\documentclass[amsmath,amsfonts,amssymb,twocolumn,showpacs,nofootinbib]{revtex4-1}
\usepackage{graphicx}
\usepackage[usenames]{color}
\usepackage{epstopdf}
\usepackage{verbatim}
\usepackage{hyperref}
\usepackage[table]{xcolor}

\newcommand{\pd}{\partial}
\newcommand{\extd}{\mathrm{d}}

\newcommand{\tbnd}{T\mathcal{M}}
\newcommand{\cotbnd}{T^*\!\mathcal{M}}

\begin{document}


\title{The Geometry of Hamiltonian Monte Carlo}
\author{Michael Betancourt}
\affiliation{MIT Department of Physics and Laboratory for Nuclear Science, Cambridge, MA 02139, USA}
\author{Leo C. Stein}
\email{leostein@mit.edu}
\affiliation{MIT Department of Physics and Kavli Institute, Cambridge, MA 02139, USA}

\date{\today}

\begin{abstract}
With its systematic exploration of probability distributions, Hamiltonian Monte Carlo is a potent Markov
Chain Monte Carlo technique; it is an approach, however, ultimately contingent
on the choice of a suitable Hamiltonian function.  By examining both the symplectic
geometry underlying Hamiltonian dynamics and the requirements of Markov Chain
Monte Carlo, we construct the general form of admissible Hamiltonians and propose
a particular choice with potential application in Bayesian inference.
\end{abstract}

\maketitle


Since its introduction by Duane, et~al.~\cite{Duane1987} and development by Neal~\cite{Neal2011}, Hamiltonian
Monte Carlo (HMC) has proven to be a powerful Markov Chain Monte Carlo
methodology.  By utilizing Hamiltonian dynamics as a Markov transition
kernel, HMC coherently explores the space of a
target distribution and results in rapidly mixing Markov chains.  Such
dynamics, however, are dependent on a particular Hamiltonian function
and existing applications have been limited mostly to forms common in
the study of physical systems.

The feature of Hamiltonian dynamics that furnishes such efficient
Markov chains is a consequence of not any particular Hamiltonian
function, but rather an implicit symplectic geometry on the underlying
parameter space.  By first understanding the properties of this
geometry, its relation to probability measures, and then the constraints
of a well-posed transition kernel,
we determine the most general form of HMC and
discuss immediate extensions to current work.

\section*{Symplectic Geometry}

Symplectic geometry is the study of an intrinsic structure of certain differentiable
manifolds having applications spanning the physical sciences~\cite{Schutz1980}.
After introducing the tangent and cotangent bundles that derive from
a given base manifold, we show
how the latter is naturally equipped with a symplectic geometry, and
how that structure admits a flow along the cotangent bundle.

\subsection*{The Tangent and Cotangent Bundles}

A smooth, $n$-dimensional manifold $\mathcal{M}$ is defined as a
topological space where the local neighborhood of any given point
$P\in\mathcal{M}$ `looks like' $\mathbb{R}^n$.  Formally, a homeomorphism
$\psi_{\alpha}:O_{\alpha}\to U_{\alpha}$ must exist from any open subset
$O_{\alpha}\subset\mathcal{M}$ to an open subset $U_{\alpha} \subset \mathbb{R}^{n}$
such that the map $\psi_{\beta}\circ
\psi_{\alpha}^{-1}$ is continuously differentiable for any overlapping subsets
$O_{\alpha}\cap O_{\beta} \neq \emptyset$.

Of the many possible differentiable mappings between differentiable
manifolds, two of the most important are curves, which map open sets of
$\mathbb{R}$ to neighboring points on the manifold, and functions, which map each
point $P$ to $\mathbb{R}$.  The collection of
all smooth curves and all smooth functions on $\mathcal{M}$ form
differentiable manifolds in their own right, and these two manifolds
naturally equip each point $P \in \mathcal{M}$ with two local vector
spaces.  Tangent vectors of all curves passing through $P$ form an
$n$-dimensional vector space known as the tangent space, $T_{P}$; the
gradients of all manifold functions in a neighborhood of point $P$
span another $n$-dimensional vector space denoted the cotangent space,
$T^{*}_{P}$.  These two vector spaces are dual in the sense that an
element of $T_{P}$ serves as a linear transformation mapping an
element of $T^{*}_{P}$ to $\mathbb{R}$, and vice versa.

A set of $n$ functions, $\left\{ q^i\!\left( P \right)
\right\}$,\footnote{Note that the index here, $i \in \left\{ 1, \ldots,
    n \right\}$, is simply a label and need not, for example, be
  considered the components of a vector.} are coordinates on the
manifold if they uniquely identify each point, and any choice of
coordinate functions induces natural bases for the tangent and
cotangent spaces across $\mathcal{M}$: the directional derivative vectors,
$\vec{e}_i \equiv \frac{\pd}{\pd q^i}$, and gradients, $\widetilde{\extd q}{}^i$,
respectively.  These coordinate bases provide an expansion of any
element $\vec{v} \in T_{P}$ or $\tilde{p} \in T^{*}_{P}$,
\begin{align*}
\vec{v} &= \sum_{i=1}^n v^i \vec{e}_i \\
\tilde{p} &= \sum_{i=1}^n p_i \widetilde{\extd q}{}^i\,,
\end{align*}
where the components $v^{i}$ (resp. $p_{i}$) are manifold functions
depending on $\vec{v}$ (resp. $\tilde{p}$) and the $q^{i}$.  Note that the
components uniquely identify each element of the vector spaces: when
$T_{P}$ and $T^{*}_{P}$ are treated as manifolds, the component
functions $\left\{v^{i}\right\}$ (resp. $\left\{p_{i}\right\}$) serve as
proper coordinate functions on $T_{P}$ (resp. $T^{*}_{P}$).

Collecting all of the tangent spaces across the manifold yields
another $2n$-dimensional manifold called the \textit{tangent bundle},
$\tbnd$. A point in $\tbnd$ corresponds to a specific vector in a
particular vector space $T_P$ and locally the bundle factors into a
trivial product space of the base manifold $\mathcal{M}$ and the
tangent space $T_{P}$.  Coordinates on $\tbnd$ then decompose into the
direct sum of coordinates on the two manifolds, $\{q^i, v^j \}$.
Likewise the cotangent spaces can be collected into the cotangent
bundle $\cotbnd$ where each point is identified by the coordinates
$\{q^i, p_j\}$.  In analogy with physical systems, the two bundles
$\tbnd$ and $\cotbnd$ are also known as configuration space and phase
space, respectively (Table \ref{tab:bundleNames}).

\begin{table*}
\begin{center}
	\rowcolors{2}{}{gray!35}
	\renewcommand{\arraystretch}{1.2}
	\begin{tabular}{c@{}c@{}c@{}c}
		\textbf{Manifold} &
    \begin{minipage}[t]{2cm}
    \textbf{Manifold Name}
    \vspace{0.5\baselineskip}
    \end{minipage}
    &
    \begin{minipage}[t]{2cm}
    \textbf{Coordinate Functions}
    \end{minipage}&
    \begin{minipage}[t]{2cm}
    \textbf{Coordinate Names}
    \end{minipage}
    \\
    $\mathcal{M}$ & Position Space & $q^i$ & Position \\
		$\tbnd$ & \;\;Configuration Space \;\;& $q^i, v^j$ & Position, Velocity \\
		$\cotbnd$ & Phase Space & $q^i, p_j$ & \;\; Position, Momenta \;\; \\
	\end{tabular}
	\caption{The tangent and cotangent bundles, and their coordinates, are often referenced in analogy to physical systems in classical mechanics.}
	\label{tab:bundleNames}
\end{center}
\end{table*}

\subsection*{The Symplectic Form}

A change of coordinates on the base manifold $\mathcal{M}$,
\begin{equation*}
\left\{ q^i \right\} \rightarrow \left\{ Q^{i} \right\},
\end{equation*}
induces a unique linear transformation on the basis one-forms $\widetilde{\extd q}{}^i$ on
$\mathcal{M}$,
\begin{equation}
\label{eq:dqtransf}
\widetilde{\extd Q}{}^{i} = \sum_{j=1}^{n} \frac{\pd Q^{i}}{\pd
  q^j} \widetilde{\extd q}{}^j\,,
\end{equation}
and on the coordinate decomposition of a one-form $\tilde{p}$,
\begin{equation}
\label{eq:ptransf}
P_{i} = \sum_{j=1}^{n} \frac{\pd q^j}{\pd Q^{i}} p_j\,.
\end{equation}
Note that the Jacobian and inverse Jacobian of the coordinate
transformation, $\pd q^{i}/\pd Q^{j}$ and
$\pd Q^{i}/\pd q^{j}$ respectively, are matrix inverses,
\begin{equation*}
\sum_{j=1}^{n}\frac{\pd Q^{i}}{\pd q^j} \frac{ \pd q^{j} }{ \pd Q^{k} } = \delta^{i}{}_{k}\,.
\end{equation*}

Coordinate transformations on $\mathcal{M}$ also induce a family of transformations
of the fields on $\cotbnd$; these \textit{point transformations}
do not mix the $\left\{ q^{i} \right\}$ and $\left\{ p_{i} \right\}$
coordinates and preserve the trivial fiber structure of the manifold.
Specifically, the one-forms $\extd q^i$ over
$\cotbnd$ (which are just the \emph{vertical lifts} of the one-form
fields $\widetilde{\extd q}{}^i$ over $\mathcal{M}$) transform as in
Eq.~\eqref{eq:dqtransf}, while the coordinate functions $p_i$
 transform as in Eq.~\eqref{eq:ptransf}. All of the
coordinate functions, $\left\{ q^i, p_j \right\}$, and their associated basis
one-forms $\left\{ \extd q^i, \extd p_j \right\}$ are clearly
dependent on the coordinates of the base manifold.

Because the two transformations in
Eq.~\eqref{eq:dqtransf} and Eq.~\eqref{eq:ptransf} are inverses,
however, there are some natural objects on $\cotbnd$ independent of the manifold coordinates.
In particular, the one-form $\theta$,\footnote{The resemblance of $\theta$
and an expansion of a one-form $\tilde{p}$ on $\mathcal{M}$ is a result
of ambiguous notation.  Here $p_i$ are coordinate functions on $\cotbnd$,
and $\extd q^i$ are one-forms on $\cotbnd$. This one-form
$\theta$ (or its negation $-\theta$) is sometimes called the
tautological form.}
\begin{equation*}
\theta \equiv \sum_{i=1}^n -p_i \extd q^i\,.
\end{equation*}
is coordinate independent by construction. The
exterior derivative $d\theta$,
\begin{equation}
\label{eq:sympldef}
\omega \equiv \extd\theta = \sum_i \extd q^i \wedge \extd p_i\,,
\end{equation}
is more subtle: although the transformation of $\extd p_i$ involves derivatives of the
Jacobian, $\omega$ is coordinate independent,
\begin{equation*}
\omega' = \sum_{j} \extd Q^{j} \wedge \extd P_{j} = \sum_i \extd q^i \wedge \extd p_i = \omega\,.
\end{equation*}
This can be seen either by explicit calculation~\cite{Schutz1980},
or more simply by noting that,
because $\theta$ is coordinate independent, the exterior derivative
$\extd\theta$ must be as well.

This \textit{symplectic form}, $\omega$, is
a two-form, or anti-symmetric tensor of rank $\left(0, 2\right)$,
which maps a given vector field $\vec{X}$ to a one-form
$\tilde{X}$,
\begin{equation*}
\tilde{X}\equiv \omega(\vec{X},\cdot)\,.
\end{equation*}
Because $\omega$ is non-degenerate, this linear transformation is
invertible, and any one-form $\tilde{Y}$ over phase space also
defines a unique vector field $\vec{Y}$ such that
$\omega (\vec{Y},\cdot) \equiv \tilde{Y}$.  Starting from any function
$H \! \left( \mathbf{q}, \mathbf{p} \right)$ on $\cotbnd$ one can
construct a one-form via the exterior derivative, $\extd H$,
and, from this, define the \emph{Hamiltonian vector field}
$\vec{X}_{H}$ associated with $H$ such that $\omega(\vec{X}_{H},\cdot)
= \extd H$.

\subsection*{Hamiltonian Flow}

By specifying a direction at each point $P \in \cotbnd$, the vector field
$\vec{X}_{H}$ defines curves through the manifold. These curves
cover the entire manifold without intersecting, mapping
every point $P$ to another along the local curve and
generating a \textit{flow} of the entire cotangent bundle.

In particular, the vector field differentiates any function $f$ along
the defined vector at each point, $\vec{X}_{H}(f) = \left( \vec{X}_{H}
\right)^i \frac{\pd}{\pd x^i}(f)$, where $x^i = \left\{ q^{i}, p_{j} \right\}$ are the coordinates of
$\cotbnd$. If the curves are parametrized by $t \in \mathbb{R}$,
then the action of the vector field is simply $\vec{X}_{H}(f) =
\frac{d}{dt}(f)$.  Applying the vector field $\vec{X}_{H}$ to the
coordinates functions themselves yields ordinary differential equations of each
coordinate along the integral curves,
\begin{align*}
\dot{q}^i \! \left( \mathbf{q}, \mathbf{p} \right) &\equiv
\vec{X}_{H}(q^i) = \frac{\extd q^i}{\extd t} = +\frac{ \partial H }{ \partial p_i }\\
\dot{p}_i \! \left( \mathbf{q}, \mathbf{p} \right) &\equiv
\vec{X}_{H}(p_i) = \frac{\extd p_i}{\extd t} = -\frac{ \partial H }{ \partial q^i }\,,
\end{align*}
known as \textit{Hamilton's equations} in classical
mechanics.\footnote{The resulting Hamiltonian flow is often denoted
  Hamiltonian evolution, or Hamiltonian dynamics.}${}^{,}$\footnote{Local
  Hamiltonian flow is not always adequate: constraints of the
  coordinate functions, for example, require discontinuous jumps in
  momentum otherwise known as specular reflection
  (Appendix~\hyperref[sec:specrefl]{A}).}

Similarly, the rate of change of an arbitrary scalar function, $f$, along the
integral curves of $\vec{X}_{H}$ is given by
\begin{align*}
\frac{\pd f}{\pd t} &=\vec{X}_{H}(f) \\
&= \extd f(\vec{X}_{H}) \\
&= \omega( \vec{X}_{f}, \vec{X}_{H} ) \\
&= \sum_i \frac{\pd f}{\pd q^i}\frac{\pd H}{\pd p_i} - \frac{\pd f}{\pd p_i}\frac{\pd H}{\pd q^i} \\
&\equiv \left\{ f, H \right\} \, ,
\end{align*}
where $\left\{ f, H \right\}$ is known as the Poisson bracket.  Note
that the function $f$ is invariant along integral curves if the
Poisson bracket vanishes; in particular the initial function $H$, or
Hamiltonian, is always preserved,
\begin{equation*}
\frac{\pd H}{\pd t} = \vec{X}_{H}(H) = \left\{H, H \right\} = 0.
\end{equation*}

The rate of change of a general geometric object along the integral
curves of $\vec{X}_{H}$ is given by using the Lie derivative,
$\mathcal{L}_{\vec{X}_{H}}$.\footnote{The scalar functions considered
  above are geometric objects in their own right and, indeed, the Lie
  derivative of a scalar field agrees with the action of the vector
  field.} By definition the symplectic form is closed,
$\extd\omega = 0$, and as a result the Lie derivative along $\vec{X}_{H}$ of $\omega$
vanishes,
\begin{equation*}
\mathcal{L}_{\vec{X}_{H}}\omega = 0 \,.
\end{equation*}
Consequently, the Lie derivative of the top-rank differential volume
form $\Omega$
\begin{equation*}
\Omega \equiv \omega^{n} \equiv \underbrace{\omega \wedge \omega \wedge \ldots \wedge \omega}_{n\mathrm{~times}}
\end{equation*}
also vanishes,
\begin{equation*}
\mathcal{L}_{\vec{X}_{H}}\Omega = 0\,,
\end{equation*}
implying that differential volume elements of phase space are
preserved under the evolution along the integral curves.  

Note that, with both the Hamiltonian and the differential phase space
volume preserved, the phase space density $F(H)\Omega$ for any smooth
scalar function $F:\mathbb{R}\to\mathbb{R}$ is also invariant along
integral curves,\footnote{This geometric statement is equivalent to
  Liouville's theorem in statistical mechanics~\cite{Kardar2007}.}
\begin{align*}
\mathcal{L}_{\vec{X}_{H}} F(H)\Omega 
&= \left(\mathcal{L}_{\vec{X}_{H}} F(H)\right)\Omega + F(H)
\mathcal{L}_{\vec{X}_{H}} \Omega \\
&= F'(H) \left(\mathcal{L}_{\vec{X}_{H}} H\right)\Omega\\
&= 0\,.
\end{align*}
This property is inherent to the symplectic geometry of the cotangent
bundle and holds for any choice of Hamiltonian and scalar function
$F$.


\section*{The Identification of Forms and Measures}

A top rank form on a $n$-dimensional manifold is a tensor, independent
of the manifold coordinates.  Given a particular choice of coordinates $\left\{ x^{i} \right\}$
defined across the entire manifold, however, any such top-rank form can be decomposed as
\begin{equation*}
\mbox{\boldmath{$\phi$}}= f_{\mathbf{x}} \!\left( \mathbf{x} \right) \mathrm{d}^{n} \mathbf{x} \, ,
\end{equation*}
where $ f_{\mathbf{x}} \!\left( \mathbf{x} \right)$ is a scalar function determined
by the coordinates and $\mathrm{d}^{n} \mathbf{x}$ is the volume form of the
coordinates constructed by wedging the gradients of each coordinate function
together,
\begin{equation*}
\mathrm{d}^{n} \mathbf{x} = \mathrm{d} x^{1} \wedge \cdots \wedge \mathrm{d} x^{n}.
\end{equation*}
Introducing the object $f$ which evaluates to the
appropriate $f_{\mathbf{x}} \! \left( \mathbf{x} \right)$ for any given set of coordinate functions 
$\left\{x^{i} \right\}$, the decomposition becomes
\begin{equation*}
\mbox{\boldmath{$\phi$}}= f \! \left( \mathbf{x} \right) \mathrm{d}^{n} \mathbf{x} \, .
\end{equation*}

By construction, the objects $f \! \left( \mathbf{x} \right)$ and 
$\mathrm{d}^{n} \mathbf{x}$ in the decomposition depend 
on the coordinates: under a change of coordinates from 
$\left\{x^{i} \right\} \rightarrow \left\{X^{i} \right\}$ the two terms 
transform by acquiring a determinant of the Jacobian matrix $\pd \mathbf{x} /\pd \mathbf{X}$,
\begin{align*}
f \!\left( \mathbf{X} \right) &= \left| \frac{ \partial \mathbf{x} }{ \partial \mathbf{X} } \right|^{+1} f \!\left( \mathbf{x} \right) \\
\mathrm{d}^{n} \mathbf{X} &= \left| \frac{ \partial \mathbf{x} }{ \partial \mathbf{X} } \right|^{-1} \mathrm{d}^{n} \mathbf{x}\,.
\end{align*}
These objects are not tensors but rather \textit{tensor densities}. In
general a tensor density of weight $w$ transforms as\footnote{The
definition of the sign of the weight can vary within the literature;
here we use the convention of \cite{Schutz1980}.}
\begin{equation*}
\rho \! \left( \mathbf{X} \right) = \left| \frac{ \partial \mathbf{x} }{ \partial \mathbf{X} } \right|^{w} \rho \! \left( \mathbf{x} \right),
\end{equation*}
with $f\! \left( \mathbf{x} \right)$ and $\mathrm{d}^{n} \mathbf{x}$
immediately identified as tensor densities of weight $+1$ and $-1$,
respectively.  Note that, when the two objects are combined, the additional
Jacobian factors cancel to give a coordinate independent object as
expected of the original tensor.

The topological space that forms the manifold $\mathcal{M}$ can also be endowed with
a coordinate independent measure that, given coordinates, can
be decomposed into two coordinate-dependent objects~\cite{Folland1999}.
Any Borel measure $\mu$ can be written as
\begin{equation*}
\mathrm{d} \mu = f \! \left( \mathbf{x} \right) \mathrm{d}^{n} \mathbf{x},
 \end{equation*}
where $\mathrm{d}^{n} \mathbf{x} $ is now the Lebesgue measure on the chosen
coordinates and $f \! \left( \mathbf{x} \right)$ is the Radon-Nikodym derivative
of $\mu$ with respect to $\mathrm{d}^{n} \mathbf{x}$.  Under a coordinate
transformation, these two objects acquire a factor of the Jacobian just as above,
\begin{align*}
f \! \left( \mathbf{X} \right) &= \left| \frac{ \partial \mathbf{x} }{ \partial \mathbf{X} } \right|^{+1} \! f \left( \mathbf{x} \right) \\
\mathrm{d}^{n} \mathbf{X} &= \left| \frac{ \partial \mathbf{x} }{ \partial \mathbf{X} } \right|^{-1} \mathrm{d}^{n} \mathbf{x} .
\end{align*}
Indeed the similarity of the two systems is no coincidence: provided
that the manifold can be oriented, the Riesz representation theorem
guarantees that the space of top-rank forms is locally isomorphic to
the space of measures \cite{Folland1999}, providing an identification
between the form \mbox{\boldmath{$\phi$}} and measure $\mu$ as well as
the terms in their above decompositions.  If the form is everywhere
positive and the integral over all of $\mathcal{M}$ is finite then the
corresponding measure becomes a normalizable Borel measure
from which one can build a probabilistic space.

Note that objects on the cotangent bundle can be densities with respect
to coordinate transformations on $\cotbnd$, acquiring factors of
$\left| \pd \left( \mathbf{q}, \mathbf{p} \right) /\pd \left(
    \mathbf{Q}, \mathbf{P} \right) \right|$, or with respect to
transformations on $\mathcal{M}$, picking up only factors of $\left|
  \pd\mathbf{q}/\pd\mathbf{Q} \right|$.  The point transformations
introduced above are a special case of \textit{symplectomorphisms},
which preserve the symplectic form $\omega$ and have Jacobian determinant
$\left| \pd \left( \mathbf{q}, \mathbf{p} \right) /\pd \left(
    \mathbf{Q}, \mathbf{P} \right) \right|=+1$. Therefore, any object
transforming as a density with respect to $\cotbnd$, such as those
in the above decompositon, are actually invariant under transformations
that preserve the structure of the cotangent bundle.


\section*{Integral Curves As Markov Transition Kernels}

Because the volume form $\Omega$ is nowhere zero, a symplectic manifold
can always be oriented and, consequently, the form $F \left(H \right) \Omega$
can always be identified with a measure.  Moreover, if $F$ is restricted
to positive, integrable functions,
\begin{equation*}
F: \mathbb{R} \rightarrow \mathbb{R}^{+}\,
\; \mathrm{s. t.} \;
\int_{\mathcal{M}} F \left( H \right) \Omega \in \mathbb{R}^{+}\,,
\end{equation*}
then the corresponding measure will be a Borel measure.

Taking $F \left( H \right) = \exp\left(C -H \right)$,\footnote{Other choices for $F$ are possible, 
but this exponential form has a computational advantage when considering the
ubiquitous distributions from the exponential family.} a given Hamiltonian uniquely defines
a probability measure
\begin{align*}
\mathrm{d} \mu &= \pi \! \left( \mathbf{q}, \mathbf{p} \right) d\mathbf{q} d\mathbf{p} \\
&= \exp \left[ C - H \! \left( \mathbf{q}, \mathbf{p} \right)
\right] \Omega,
\end{align*}
where $C \in \mathbb{R}$ is determined by the normalization.
Hamiltonian flow maps $\mu$ into itself and thus defines a proper 
Markov transition kernel for the distribution
$\pi \! \left( \mathbf{q}, \mathbf{p} \right)$. Note
that, in practical applications the dynamics must be performed
numerically and the measure will not be conserved
exactly~\cite{Leimkuhler2004}. Any subsequent bias, however, can be
avoided by treating the flow as a Metropolis proposal function instead
of the transition itself \cite{Duane1987, Neal2011}.

Now if the gradients of the Hamiltonian satisfy
\begin{align*}
\dot{q}^i \! \left( \mathbf{q}, -\mathbf{p} \right) &= -\dot{q}^i \! \left( \mathbf{q}, \mathbf{p} \right) \\
\dot{p}_i \! \left( \mathbf{q}, -\mathbf{p} \right) &= +\dot{p}_i \! \left( \mathbf{q}, \mathbf{p} \right)
\end{align*}
then the resulting flow is reversible: reflecting the momenta and
evolving the same distance along the integral curves recovers the
initial configuration.  By adding such a reflection to the end of each
evolution, the resulting transition has detailed balance with respect
to $\pi \! \left( \mathbf{q}, \mathbf{p} \right)$, which then becomes the
unique stationary distribution.

The augmented transitions, however, do not produce a convergent Markov
chain because the evolution explores only level sets of probability
density,
\begin{equation*}
\pi \! \left( \mathbf{q}, \mathbf{p} \right) = \pi \! \left( \mathbf{q}_{0}, \mathbf{p}_{0} \right) \propto \exp \left( - H \! \left( \mathbf{q}_{0}, \mathbf{p}_{0} \right) \right).
\end{equation*}
Movement across the probability contours can be introduced by adopting
a Gibbs strategy and alternating each Hamiltonian transition with
transitions from any conditional distribution spanning the contours.
The combined transitions guarantee ergodicity and, provided that the
transitions are tuned to avoid cycles in phase space and subsequent
periodicity, the resulting Markov chain will converge to $\pi \! \left(
  \mathbf{q}, \mathbf{p} \right)$.

While any reversible Hamiltonian defines both a probability
distribution and respective Markov chain, practical applications
demand the reverse construction: what Hamiltonians, and hence
well-behaved Markov chains, are consistent with a given target
distribution $\pi \! \left(\mathbf{x} \right)$?

Provided a decomposition of
the random variables into positions and momenta, $\mathbf{x} = \left\{
  \mathbf{q}, \mathbf{p} \right\}$, the support of the target
distribution could be identified with the full cotangent bundle and $\pi \!
\left( \mathbf{x} = \left\{ \mathbf{q}, \mathbf{p} \right\} \right)$
would fully define a Hamiltonian.  There is no natural motivation for 
such a decomposition, however, and one
resulting in a reversible Hamiltonian system need not even
exist.\footnote{A trivial example being a distribution defined on an
  odd-dimensional space.}

If we appeal to the factorization of the cotangent bundle into
$\left\{q^{i}\right\}$ and $\left\{p_{i}\right\}$ coordinates and instead identify
the random variables with the position manifold $\mathcal{M}$, however, 
then $\mathbf{x} = \mathbf{q}$ and
the target distribution constrains only the marginal distribution of
$\pi \!\left( \mathbf{q}, \mathbf{p} \right)$,
\begin{equation*}
\pi \! \left( \mathbf{x} \right) = \left. \int \mathrm{d}^{n} \mathbf{p} \, \pi \! \left( \mathbf{q}, \mathbf{p} \right) \right|_{ \mathbf{q} = \mathbf{x} }.
\end{equation*}
Any joint distribution defined over the entire cotangent bundle
factors,
\begin{equation*}
\pi \! \left( \mathbf{q}, \mathbf{p} \right) = \pi \! \left( \mathbf{p} | \mathbf{q} \right) \pi \! \left( \mathbf{q} \right),
\end{equation*}
and the residual freedom in the choice of conditional distribution, $\pi \!
\left( \mathbf{p} | \mathbf{q} \right)$, can be engineered to not only
guarantee reversibility but also be easily sampled to ensure that the
entire procedure remains computationally efficient.  Once the latent
variables $\mathbf{p}$ are marginalized out, the Markov chain generates the
desired samples from $\pi \! \left( \mathbf{q} \right)$.

This final identification completes the construction of a general MCMC
procedure: a given target distribution $\pi \! \left( \mathbf{q} \right)$
and the selection of an appropriate conditional distribution, $\pi \!
\left( \mathbf{p} | \mathbf{q} \right)$, defines Hamiltonian dynamics
and a well-behaved transition kernel.  Because the dynamics
incorporate gradients of the target distribution, the transitions
systematically explore the target distribution and avoid the random
walk behavior typical of other MCMC techniques~\cite{Neal2011}.
Moreover, the freedom in the conditional distribution offers the
potential for including more information about the target distribution
and, consequently, further improving the performance of the resulting
Markov chain.


\section*{Admissible Hamiltonians}

With the above considerations, the most general Hamiltonian yielding
the desired Markov chain is
\begin{align*}
H &= - \log \pi \! \left( \mathbf{q}, \mathbf{p} \right) + C \\
&= - \log \pi \! \left( \mathbf{p} | \mathbf{q} \right) \pi \! \left( \mathbf{q} \right) + C \\
&=- \log \pi \! \left( \mathbf{p} | \mathbf{q} \right) - \log \pi \! \left( \mathbf{q} \right)  + C \\
& \equiv T \! \left( \mathbf{q}, \mathbf{p} \right) + V \! \left( \mathbf{q} \right) + C.
\end{align*}
Note that $T$ and $V$ are neither scalars nor scalar
densities. While the joint distribution $\pi \! \left( \mathbf{q}, \mathbf{p} \right)$ is
invariant under point transformations (and indeed all symplectomorphisms), the distributions in the
decomposition are densities of opposite weight with respect to
coordinate transformations on the position manifold $\mathcal{M}$,
\begin{align*}
\pi \! \left( \mathbf{P} | \mathbf{Q} \right) &= \left| \frac{ \partial \mathbf{q} }{ \partial \mathbf{Q} } \right|^{-1} \pi \! \left( \mathbf{p} | \mathbf{q} \right) \\
\pi \! \left( \mathbf{Q} \right) &= \left| \frac{ \partial \mathbf{q} }{ \partial \mathbf{Q} } \right|^{+1} \pi \! \left( \mathbf{q} \right)\,,
\end{align*}
and the terms in the Hamiltonian must transform as
\begin{align}
\label{eq:tTrans}
T \! \left( \mathbf{Q}, \mathbf{P} \right) &= T \! \left( \mathbf{q}, \mathbf{p} \right)  + \log \left| \frac{ \partial \mathbf{q} }{ \partial \mathbf{Q} } \right| \\
 V \! \left( \mathbf{Q} \right) &=  V \! \left( \mathbf{q} \right) - \log \left| \frac{ \partial \mathbf{q} }{ \partial \mathbf{Q} } \right|.
\end{align}
When added together the additional factors cancel to yield the scalar Hamiltonian. 

While the potential energy $V \! \left( \mathbf{q} \right)$ is fully
specified by the target distribution, the kinetic energy is
constrained by only the defining normalization of a conditional
distribution
\begin{equation*}
\int \mathrm{d}^{n} \mathbf{p} \, \exp \left( - T \! \left( \mathbf{q},
    \mathbf{p} \right) \right) \in \mathbb{R}^{+},
\end{equation*}
i.e.~a finite positive number independent of $\mathbf{q}$,
and the demands of detailed balance,
\begin{align*}
T \! \left( \mathbf{q}, -\mathbf{p} \right) &= T \! \left( \mathbf{q}, \mathbf{p} \right) \\
\frac{ \partial T }{ \partial p_i } \! \left( \mathbf{q}, - \mathbf{p} \right) &= - \frac{ \partial T }{ \partial p_i } \! \left( \mathbf{q}, \mathbf{p} \right).
\end{align*}

Reversibility is assured for any kinetic energy of the form
\begin{equation*}
T \! \left( \mathbf{q}, \mathbf{p} \right) = \tau_{1} \! \left( \mathbf{q}, \mathbf{p} \right) + \tau_{1} \! \left( \mathbf{q}, - \mathbf{p} \right) + \tau_{2}  \! \left( \mathbf{q} \right),
\end{equation*}
where the first two terms can, in general, be decomposed into a sum of
completely symmetric tensors 
$T_{(n)}$ of type $\left( n, 0\right)$ with $n$ even and positive,
\footnote{In order to satisfy the defining
transformation property (\ref{eq:tTrans}), the contribution from 
$\tau_{1} \! \left( \mathbf{q}, \mathbf{p} \right) + \tau_{1} \! \left(
  \mathbf{q}, - \mathbf{p} \right)$ must be a scalar and, consequently,
the $T_{(n)}$ must be tensors.  The second term $\tau_{2}  \! \left( \mathbf{q} \right)$ is
ultimately responsible for the introduction of the necessary 
factor of $\log \left| \partial \mathbf{q} / \partial \mathbf{Q} \right|$.}
\begin{equation*}
\tau_{1} \! \left( \mathbf{q}, \mathbf{p} \right) + \tau_{1} \! \left(
  \mathbf{q}, - \mathbf{p} \right) =
\sum_{n =2,4,\ldots}\sum_{j_1\ldots j_n}
 p_{j_{1}} \ldots p_{j_{n}} T_{(n)}^{j_{1} \ldots j_{n} } \!\left( \mathbf{q} \right).
\end{equation*}

A particularly useful choice is any scalar function of a non-degenerate
quadratic form in the momenta,
\begin{equation*}
T \! \left( \mathbf{q}, \mathbf{p} \right) = \tau \!\left( \sum_{ij} p_{i}
  p_{j} A^{ij} \! \left( \mathbf{q} \right) \right) + t_{2} \! \left( \mathbf{q} \right).
\end{equation*}
Here the tensor $\mathbf{A} \! \left( \mathbf{q} \right)$, or more
appropriately its inverse, effectively
serves as a spatially-dependent linear transformation of the coordinates $\mathbf{q}$ ---
if the transformation simplifies the structure of the target distribution
then the resulting Markov chain promises to explore the space much more efficiently.
In particular, the incorporation of the derivatives of the potential
can help to locally standardize the distribution, avoiding
complications due to the narrow valleys characteristic of strong correlations.

Taking $\tau$ to be the identity gives
\begin{equation} \label{eqn:quadT}
T \! \left( \mathbf{q}, \mathbf{p} \right) = \frac{1}{2} \sum_{ij} p_{i} p_{j} 
\Lambda^{ij}\!\left( \mathbf{q} \right) - \frac{1}{2} \log \left| \mathbf{\Lambda} \! \left( \mathbf{q} \right) \right|,
\end{equation}
which results in a gaussian conditional distribution,
\begin{equation*}
\pi \! \left( \mathbf{p} | \mathbf{q} \right) = \mathcal{N} \! \left( \mathbf{p} | \mathbf{0}, \mathbf{\Lambda} \right).
\end{equation*}
Notice how the normalization of the conditional gaussian introduces
the determinant of $\mathbf{\Lambda}$, which transforms as a tensor
density of weight $-2$ with respect to coordinate transformations on
$\mathcal{M}$.\footnote{Provided that $k + l$ is even, the determinant
  of a rank $(k, l)$ tensor exists and transforms as a scalar density
  of weight $l - k$.}  Under a point transformation the normalization
becomes
\begin{align*}
\frac{1}{2} \log \left| \mathbf{\Lambda} \! \left( \mathbf{Q} \right) \right| &= \frac{1}{2} \log  \left| \partial \mathbf{q} / \partial \mathbf{Q} \right|^{-2} \left| \mathbf{\Lambda} \! \left( \mathbf{q} \right) \right| \\
&= \frac{1}{2} \log \left| \mathbf{\Lambda} \! \left( \mathbf{q} \right) \right| + \frac{1}{2} \log \left| \partial \mathbf{q} / \partial \mathbf{Q} \right|^{-2} \\
&= \frac{1}{2} \log \left| \mathbf{\Lambda} \! \left( \mathbf{q} \right) \right| - \log \left| \partial \mathbf{q} / \partial \mathbf{Q} \right|,
\end{align*}
introducing exactly the necessary factor of the Jacobian
to ensure the proper transformation of $T$.  The identification of 
measures with forms ensures consistency between the two perspectives.

If the covariance $\mathbf{\Sigma} =
\mathbf{\Lambda}^{-1}$ is proportional to the identity then
the Hamiltonian reduces to that of classical mechanics and the form
used in the first implementations of HMC~\cite{Duane1987}.  A
nontrivial but constant covariance matrix allows for a global
rescaling and rotation of the target distribution~\cite{Neal2011}, and
a spatially dependent covariance transforms the target distribution
locally~\cite{Girolami2011}.

Because integrability requires it to be positive-definite and
non-degenerate, the spatially-varying covariance can also be interpreted as
a Riemannian metric and, from this perspective, the Hamiltonian
evolution locally parallels geodesics of a curved
manifold~\cite{Calin2004}.  This additional geometric structure
embedded within the symplectic geometry creates the potential to
further improve performance with the application of more tools from
differential geometry.  Utilizing any such possibility, however, first
requires a choice of the spatially-varying
covariance.

\subsection*{Choices of the Covariance Matrix}

The Fisher-Rao metric ubiquitous in information geometry,
\begin{equation*}
\Sigma_{ij} = \mathbb{E}_{\mathbf{y}} \left[ \frac{ \partial \log \pi \! \left( \mathbf{x} | \mathbf{y} \right) }{ \partial x^{i} } \frac{ \partial \log \pi \!\left( \mathbf{x} | \mathbf{y} \right) }{ \partial x^{j} }  \right],
\end{equation*}
or given particular coordinate functions
\begin{equation*}
\Sigma_{ij} = \mathbb{E}_{\mathbf{y}} \left[ \frac{ \partial V \! \left( \mathbf{x} | \mathbf{y} \right) }{ \partial x^{i} } \frac{ \partial V \! \left( \mathbf{x} | \mathbf{y} \right) }{ \partial x^{j} } \right],
\end{equation*}
provides an obvious candidate for the covariance matrix, and its use in HMC has proven
successful when the expectation can be performed analytically~\cite{Girolami2011}.
While the Fisher-Rao
metric does incorporate derivatives of the target distribution,
however, it requires integrating over the $\mathbf{y}$ to ensure
positive-definiteness.  In a Bayesian application this necessitates
expectation of the posterior over the ensemble of possible data sets;
not only is the expectation contrary to the Bayesian philosophy of
inference based solely on the measured data, it also washes out the
local structure particular to a given data set that can prove
important in posterior exploration.

The geometric perspective, however, suggests another possibility free from
the unwanted marginalization.  A ``background'' Riemannian metric
$\mbox{\boldmath{$\sigma$}}$ defined on $\mathcal{M}$
induces a metric on the graph of the potential $V$,
\footnote{The graph is an $n$-dimensional submanifold in the
$(n+1)$-dimensional manifold with coordinates
$\left(\mathbf{q}, V(\mathbf{q})\right) $.}
\begin{equation*}
\Sigma_{ij} = \sigma_{ij} (\mathbf{q}) + \frac{\pd V \! (\mathbf{q})}{\pd q^{i}}\frac{\pd V \! (\mathbf{q})}{\pd q^{j}} \, .
\end{equation*}
Unfortunately, $\mathbf{\Sigma}$ does not transform as a proper
tensor because $V$ is not a proper scalar function. The
additional structure afforded by $\mbox{\boldmath{$\sigma$}}$, however, admits
the necessary correction: because the determinant $| \mbox{\boldmath{$\sigma$}} |$ is a scalar
density of weight $+2$ with respect to transformations on $\mathcal{M}$, the quantity
\begin{equation*}
\bar{V}\! (\mathbf{q}) \equiv V\! (\mathbf{q}) + \frac{1}{2} \log \left| \mbox{\boldmath{$\sigma$}} (\mathbf{q}) \right|
\end{equation*}
is a true scalar and the modified metric
\begin{equation*}
\bar{\Sigma}_{ij} (\mathbf{q}) \equiv \sigma_{ij} (\mathbf{q}) + \frac{\pd \bar{V} \! (\mathbf{q})}{\pd q^{i}}\frac{\pd \bar{V} \! (\mathbf{q})}{\pd q^{j}}
\end{equation*}
a true tensor.

Note the similarity of the outer product  $(\pd_{i} \bar{V} )(\pd_{j} \bar{V} )$
to the argument of the expectation value in the Fisher-Rao metric: this induced
metric features a similar structure without the undesired expectation.
Moreover, $\bar{\mathbf{\Sigma}}$ is simply a rank-$1$ update of the
background geometry and the metric can be efficiently inverted with the 
Sherman-Morrison-Woodbury formula~\cite{Golub1996},
\begin{equation*}
\bar{\Lambda}^{ij} = \lambda^{ij} - \frac{(\pd^{i} \bar{V})(\pd^{j}\bar{V})}{1+(\pd^{l}\bar{V})(\pd_{l}\bar{V})}\,,
\end{equation*}
where \mbox{\boldmath{$\lambda$}} is the inverse of \mbox{\boldmath{$\sigma$}} 
(satisfying $\lambda^{ik}\sigma_{kj} = \delta^{i}{}_{j}$), and $\pd^{i} =
\lambda^{ij}\pd_{j}$.  

If \mbox{\boldmath{$\sigma$}} is homogeneous (i.e.~independent of
position) then the inversions necessary at each iteration of the Hamiltonian
evolution can be computed at order $O \left(n^{2}\right)$, significantly faster
 than the $O \left(n^{3}\right)$ required for the inversion of the dense Fisher-Rao metric.
The Christoffel coefficients of the metric $\bar{ \mathbf{\Sigma} }$,
which specify the Hamiltonian flow, are straightforward to calculate in this case,
\footnote{Because the determinant of the homogeneous metric 
$|\mbox{\boldmath{$\sigma$}}|$ is also position independent,
$\pd_{i}\bar{V} = \pd_{i}V$ and the metrics coincide,
$\bar{\mathbf{\Sigma}}=\mathbf{\Sigma}$.}

\begin{equation*}
\Gamma^{i}{}_{jk} = \frac{(\pd^{i}\, V)(\pd_{j}\pd_{k} V)}{1+(\pd^{l} V)(\pd_{l} V)}\,,
\end{equation*}
also requiring only $O(n^{2})$ operations and allowing the entire Hamiltonian flow
to be calculated at the same order.
Note the appearance of $\pd_{j}\pd_{k} V$ in the Christoffel coefficients:
while $\bar{ \mathbf{\Sigma} }$ appears to incorporate only the outer
product approximation to the Hessian of $V$, the full Hessian, and all
the information it contains, is included the evolution.

Constructed from derivatives of $\bar{ \mathbf{\Sigma} }$, the
Riemann curvature tensor and its contraction, the rank-2 Ricci tensor, offer
the potential to include higher-order derivatives and, consequently,
more information into the evolution. An explicit calculation, however, shows
that, surprisingly, only second derivatives of $V$ contribute to the
Riemann tensor. Further investigation is necessary to determine the full utility of
including these tensors into the kinetic energy.

\section*{Conclusions and Future Work}

By considering the geometry of Hamiltonian dynamics and the basic
constraints of Markov chain Monte Carlo, we have constructed the most
general approach to Hamiltonian Monte Carlo (HMC).

The simplest admissible Hamiltonian given in Eq.~\eqref{eqn:quadT} reproduces existing approaches to
Hamiltonian Monte Carlo, but knowing the most general
form begs further extensions. Girolami, et~al., for example, have
considered the kinetic energy
\begin{equation*}
T \! \left( \mathbf{q}, \mathbf{p} \right) = \frac{\nu + n}{2} \log \left( 1 + \frac{ \sum_{ij = 1}^{n} p_{i} p_{j} 
\Lambda^{ij} \! \left( \mathbf{q} \right) }{\nu} \right) - \frac{1}{2} \log \left| \mathbf{\Lambda} \! \left( \mathbf{q} \right) \right|,
\end{equation*}
which gives Student's t-distribution in place of the gaussian, but
with only mixed results.  Other choices of the kinetic energy may
dramatically improve HMC for certain distributions, and there may
still be means to improve the performance universally.

Generalizing the symplectic manifold of HMC to a Poisson manifold~\cite{Weinstein1983}
offers even more possibilities.  Without the restriction of non-degenerate forms,
the Poisson manifold can accommodate distributions with spatially varying
dimensionality and perhaps even admit trans-dimensional Monte Carlo.

Given the unexplored possibilities of this geometric perspective, the future of HMC
is promising.

\acknowledgments
We thank Katherine Deck and Ila Varma for valuable
discussions and comments on a preliminary draft of this
manuscript, and Mark Girolami for reviewing an intermediate draft.
LCS acknowledges support from MIT's Solomon Buchsbaum fund.


\section*{Appendix A: Specular Reflection For General Hamiltonians}
\label{sec:specrefl}

The explicit incorporation of inequality constraints is often advantageous,
and sometimes even necessary, in the construction of certain Markov chains.
Arbitrary inequality constraints,
\begin{equation*}
C \! \left( \mathbf{q} \right) > 0,
\end{equation*}
can be incorporated into the Hamiltonian framework with the introduction of an infinite potential~\cite{Betancourt2011},
\begin{equation*}
V \! \left( \mathbf{q} \right) = \left\{ \begin{array}{rc} \infty,& C \! \left( \mathbf{q} \right) \le 0 \\ 0,&  \mathrm{else} \end{array} \right. .
\end{equation*}
A naive implementation of Hamiltonian dynamics with such a potential,
however, immediately fails because the gradient $\extd V\!\left(
  \mathbf{q} \right)$ along the boundary is undefined.

When considering the classical mechanics of a point particle, $H = \frac{1}{2} \sum_{ij} p_{i} p_{j}
\delta^{ij} + V \! \left( \mathbf{q} \right)$, the difficulties around
the potential barrier can be avoided by appealing to the exact result:
the components of momentum perpendicular to the constraint surface
reflect while preserving the value of the Hamiltonian.  This \textit{specular
  reflection} is given by
\begin{equation} \label{eqn:spec}
\Delta \mathbf{p} = \mathbf{p}' - \mathbf{p} = - 2 \left( \mathbf{p} , \hat{ \mathbf{n} } \right) \hat{ \mathbf{n} },
\end{equation}
where the unit normal along the surface of equality ($C(\mathbf{q})=0$) is
\begin{equation*}
\hat{ \mathbf{n} } = \frac{ \extd C }{\sqrt{ \left( \extd C, \extd C \right) } }
\end{equation*}
and the inner product of one-forms is induced by the symmetric $\left(2, 0\right)$ tensor $\mbox{\boldmath{$\delta$}}$,
\begin{equation*}
\left( \mathbf{a}, \mathbf{b} \right) \equiv \sum_{ij} a_{i} b_{j} \delta^{ij} = \sum_{i} a_{i} b_{i}\,.
\end{equation*}
The reflection depends only on the direction of $\extd C$ and
not the undefined $\extd V$.
Also note that only the components of the momentum perpendicular to
level sets of $C(\mathbf{q})$ transform; those parallel to level sets
are unaffected.

Generalizing specular reflection to a general Hamiltonian system is
straightforward.  The gradient $\mathbf{n} \equiv \extd C$ defines a
unique one-form at the surface of constraint equality
$C(\mathbf{q})=0$, and forms a basis with the addition of $\left(n -
  1\right)$ one-forms $\mbox{\boldmath{$\omega$}}^{i}$,
$i\in\left\{1,\ldots,n-1\right\}$, each linearly independent of
one another and $\mathbf{n}$. With this basis, the momentum
$\mathbf{p}$ at a point of reflection $\mathbf{q}$ on the constraint
boundary may be decomposed as
\begin{equation*}
\mathbf{p} = p^{\perp} \mathbf{n} + \sum_{i = 1}^{n -1} p^{\parallel}_{i} \mbox{\boldmath{$\omega$}}^{i}\,.
\end{equation*}
Because the reflection is determined entirely by the surface $C \! \left( \mathbf{q} \right) = 0$,
the momentum can transform only along the one-dimensional subspace spanned by $\mathbf{n}$.
Consequently, the $p^\parallel$ should be invariant and a general reflection is given by
\begin{equation*}
\mathbf{p} = p^{\perp} \mathbf{n} + \sum_{i = 1}^{n -1} p^{\parallel}_{i} \mbox{\boldmath{$\omega$}}^{i} \rightarrow \mathbf{p}' = \alpha p^{\perp} \mathbf{n} + \sum_{i = 1}^{n -1} p^{\parallel}_{i} \mbox{\boldmath{$\omega$}}^{i},
\end{equation*}
along with the condition that the Hamiltonian remains invariant, $H \!
\left( \mathbf{q}, \mathbf{p}' \right) = H \! \left( \mathbf{q},
  \mathbf{p} \right)$.

In the case of a quadratic kinetic energy,
\begin{equation*}
H  \!\left( \mathbf{q}, \mathbf{p} \right) = \frac{1}{2} \sum_{ij} p_{i} p_{j} \Lambda^{ij} \!\left( \mathbf{q} \right) - \frac{1}{2} \log \left| \mathbf{\Lambda} \! \left( \mathbf{q} \right) \right| + V  \!\left( \mathbf{q} \right),
\end{equation*}
conservation of the Hamiltonian requires
%
\begin{align*}
0 &= H \! \left( \mathbf{q}, \mathbf{p}' \right) - H \! \left( \mathbf{q}, \mathbf{p} \right) \\
&= \left( \alpha -1 \right) p^{\perp} \left[ \frac{1}{2} \left( \alpha + 1 \right) p^{\perp} \sum_{ij} \left( \extd C \right)_{i} \left( \extd C \right)_{j} \Lambda^{ij} \
\right.\\ &\left.\qquad\qquad\qquad\quad
+ \sum_{k = 1}^{n - 1} p^{\parallel}_{k} \sum_{ij} \left( \extd C \right)_{i} ({\omega}^k)_{j} \Lambda^{ij} \right],
\end{align*}
%
where $ \left( \extd C \right)_{i}$ and $\left(\omega^k \right)_{j}$ are the components of the respective one-forms in an arbitrary basis.

Ignoring the $\alpha = 1$ solution where the momentum is unchanged,
\begin{align*}
\alpha &= - 2 \frac{  \sum_{k = 1}^{n - 1} p^{\parallel}_{k} \sum_{ij} \left( \extd C \right)_{i} (\omega^k)_{j} \Lambda^{ij} }{ p^{\perp} \sum_{ij} \left( \extd C \right)_{i} \left( \extd C \right)_{j} \Lambda^{ij}} - 1\,,\\
\intertext{or with a bit of manipulation}
\alpha - 1 &= - 2 \frac{ \sum_{ij} \left( \extd C \right)_{i} p_{j} \Lambda^{ij} }{ p^{\perp} \sum_{ij} \left( \extd C \right)_{i} \left( \extd C \right)_{j} \Lambda^{ij} }\,,
\end{align*}
implying that the change in momentum is
\begin{align*}
\Delta \mathbf{p} &= \left( \alpha - 1 \right) p^{\perp} \mathbf{n} \\
&= - 2 \frac{ \sum_{ij} \left( \extd C \right)_{i} p_{j} \Lambda^{ij}
}{\sum_{ij} \left( \extd C \right)_{i} \left( \extd C \right)_{j}
  \Lambda^{ij} } \mathbf{n}\,.
\end{align*}
Defining an inner product induced by the tensor $\mathbf{\Lambda}$,
\begin{equation*}
\left( \mathbf{a} , \mathbf{b} \right)_{\mathbf{\Lambda}} \equiv \sum_{ij} a_{i} b_{j} \Lambda^{ij}\,,
\end{equation*}
and a unit one-form under this inner product
\begin{equation*}
\hat{ \mathbf{a} } \equiv \frac{ \mathbf{a} }{ \sqrt{ \left( \mathbf{a}, \mathbf{a} \right)_{\mathbf{\Lambda}} } }\,,
\end{equation*}
the generalized specular reflection becomes
\begin{equation} \label{eqn:genSpec}
\Delta \mathbf{p} = - 2 \left( \hat{ \mathbf{n}} , \mathbf{p} \right)_{\mathbf{\Lambda}} \hat{ \mathbf{n} }\,.
\end{equation}

The resemblance of Eq.~\eqref{eqn:genSpec} to Eq.~\eqref{eqn:spec} is welcome.  As before, the dependence of the result on only the direction of $\mathbf{n}$ ensures that the infinite
magnitude of the potential barrier provides no difficulties.  Moreover, the general result not only reduces to the classical case for $\Lambda^{ij} =
\delta^{ij}$, it also agrees with the generalization that would be
expected given the interpretation of the quadratic kinetic energy
resulting from a Riemannian (covariant) metric
$\Sigma_{ij}$ on the base manifold (where as before
$\mathbf{\Sigma}=\mathbf{\Lambda}^{-1}$).

A straightforward calculation verifies that Eq.~\eqref{eqn:genSpec} is also a
sufficient reflection solution for any kinetic energy of the form
\begin{equation*}
T \! \left( \mathbf{q}, \mathbf{p} \right) = \tau \! \left( \sum_{ij} p_{i}
  p_{j} \Lambda^{ij} \! \left( \mathbf{q} \right) \right) + \tau_{2} \!
\left( \mathbf{q} \right) \, .
\end{equation*}

\pagebreak
\bibliography{symplecticMC}

\end{document}